\documentstyle[12pt]{article}

\newlength{\dinwidth}                                                           
\newlength{\dinmargin}                                                          
\catcode`@=11

\def\@citex[#1]#2{\if@filesw\immediate\write\@auxout{\string\citation{#2}}\fi
  \def\@citea{}\@cite{\@for\@citeb:=#2\do
    {\@citea\def\@citea{,\penalty\@m}\@ifundefined
      {b@\@citeb}{{\bf ?}\@warning
       {Citation `\@citeb' on page \thepage \space undefined}}%
\hbox{\csname b@\@citeb\endcsname}}}{#1}}

\def\citer{\@ifnextchar [{\@tempswatrue\@citexr}{\@tempswafalse\@citexr[]}}

\def\@citexr[#1]#2{\if@filesw\immediate\write\@auxout{\string\citation{#2}}\fi
  \def\@citea{}\@cite{\@for\@citeb:=#2\do
    {\@citea\def\@citea{--\penalty\@m}\@ifundefined
       {b@\@citeb}{{\bf ?}\@warning
       {Citation `\@citeb' on page \thepage \space undefined}}%
\hbox{\csname b@\@citeb\endcsname}}}{#1}}
\catcode`@=12

\def\bo{{\raise.15ex\hbox{\large$\Box$}}}               
\def\face{{\raise.2ex\hbox{$\displaystyle \bigodot$}\mskip-2.2mu \llap {$\ddot
        \smile$}}}                                      

\def\leftrightarrowfill{$\mathsurround=0pt \mathord\leftarrow \mkern-6mu
        \cleaders\hbox{$\mkern-2mu \mathord- \mkern-2mu$}\hfill
        \mkern-6mu \mathord\rightarrow$}       
\def\dvec#1{\vbox{\ialign{##\crcr
        \leftrightarrowfill\crcr\noalign{\kern-1pt\nointerlineskip}
        $\hfil\displaystyle{#1}\hfil$\crcr}}}           

\def\beq{\begin{equation}}
\def\eeq{\end{equation}}

\def\beqx{\begin{displaymath}}
\def\eeqx{\end{displaymath}}

\def\beql{\begin{eqnarray}}
\def\eeql{\end{eqnarray}}

\begin{document}	

\begin{flushright}
NIKHEF/99-023\\
September 1999
\end{flushright}

\vspace{15mm}
\begin{center}
{\Large\bf\sc Klein Bottles and Simple Currents\\ }
\end{center}
\vspace{2cm}
\begin{center}
{\large L.R. Huiszoon$^\dagger$, A.N. Schellekens$^{\dagger *}$ and N. Sousa$^\dagger$}\\
\vspace{15mm}
{\it $^\dagger$NIKHEF Theory Group\\
P.O. Box 41882, 1009 DB Amsterdam, The Netherlands} \\
\vspace{7mm}
{\it $^*$University of Nijmegen\\ Nijmegen, The Netherlands}
\end{center}

\vspace{2cm}

\begin{abstract}

The standard Klein bottle coefficient in the construction of open descendants is shown to equal the Frobenius-Schur indicator of a conformal field theory. Other consistent Klein bottle projections are shown to correspond to simple currents. These observations enable us to generalize the standard open string construction from C-diagonal parent theories to include non-standard Klein bottles. Using (generalizations of) the Frobenius-Schur indicator
we prove positivity and integrality of the resulting open and closed string
state multiplicities for standard as well as non-standard Klein bottles. 

\end{abstract} 

\thispagestyle{empty}

\newpage \setcounter{page}{2}

\section{Introduction}

The construction of closed and open string vacua are not on equal footing. With the use of  conformal field theory (CFT), closed (oriented) string theories can be constructed by mass production: simply put, the problem is reduced to finding all possible modular invariants. In sharp contrast, the construction of open (unoriented) string vacua is still at a primitive level. The algoritm to construct open string theories with the use of CFT is known as the method of open descendants~\cite{sagnotti}. In short, one starts with a modular invariant closed oriented (`parent') theory, described by a torus partition function. Then one truncates the spectrum to an unoriented one by a Klein bottle projection. Generically, the resulting closed unoriented theory is inconsistent due to massless tadpoles of unphysical particles. These tadpoles are then removed by adding the right amount of open string  sectors, described by annulus and M\"obius strip. 

So the problem of constructing open string vacua can be formulated as follows: given a modular invariant parent theory, what are the allowed Klein bottle, annulus and M\"obius strip? The hope is that we can do this in a model independent way, that is, for any underlying CFT. There exists such a model independent prescription (usually referred to as the
``Cardy case") only in case of a C-diagonal\footnote{By ``C-diagonal" we mean that the torus partition
function is the charge conjugation invariant, where `C' stands for `Cardy' or `charge conjugation'. } parent theory with `standard' Klein bottle projection and symmetry preserving boundary conditions for the open strings. This prescription is used for example in~\cite{blumenhagen} \cite{comments}. In particular examples, the open descendants of non-C-diagonal invariants~\cite{blumenhagen} \cite{comments} \cite{nondiagonal} and/or `non-standard' Klein bottles~\cite{descendants}~\cite{planar} are found by use of rather lengthy and explicit computations. 

In this letter, we provide an elegant way to produce open descendants of
C-diagonal 
parents with non-standard Klein bottle projections. This construction follows from the observation that Klein bottle projections correspond to simple currents. Examples 
of this construction have been discussed  
in~\cite{descendants} \cite{properties} in a different language. But it is the language of simple currents that enables us to 
write down the model independent solution.   

In section~\ref{sec-open}, we review the method of open descendants~\cite{sagnotti} \cite{descendants} \cite{properties} with the use of partition functions. This review is not meant as a pedagogical introduction to the subject; we merely fix our notation and present the formulas that are relevant for the discussion in this letter. In the end of section~\ref{sec-open}, we discuss the C-diagonal case with a standard Klein bottle. We also provide a proof that this Klein bottle projection is consistent, by showing that the Klein bottle coefficient is equal to the Frobenius-Schur indicator of a conformal field theory. 

In section~\ref{sec-simple}, we sum up some facts about simple currents~\cite{simple}. It will turn out that the ideas of this letter are most easily discussed when dealing with order $2$ simple currents that have (half-)integer spin. In subsection~\ref{sec-simple2}, we restrict ourselves to these simple currents, and show that we can associate consistent  Klein bottle projection with them. From this observation, it is straightforward to generalize the C-diagonal case to include non-standard Klein bottles. In subsection~\ref{sec-simpleN}, we generalize this construction for order $N$ simple currents.  

Finally, in appendix~\ref{sec-form}, we derive some useful formulas involving simple currents and  the matrix $P$ and the tensor $Y$, which are encountered frequently in open string constructions.

\section{A brief review of open descendants} \label{sec-open}

The starting point in the construction of open string vacua is a consistent closed oriented string theory, whose spectrum is encoded in a modular invariant  torus partition function  
\begin{equation} \label{eq:torus}
		T = \sum_{ij} \chi_i  Z_{ij} \bar{\chi}_{j} \;\; ,
\end{equation}
where $Z$ is a (symmetric) modular invariant. The index $i$ labels the (chiral) primaries and $\chi_i$ are the characters. We assume the CFT to be rational, so the range of this index is finite. 

The second step is the construction of a closed unoriented theory out of the closed oriented theory given by equation~(\ref{eq:torus}). This is done by halving the torus and adding to it a (halved) Klein bottle 
\begin{equation}
	K = \frac{1}{2} \sum_i K^i \chi_i \;\; .
\end{equation}
The Klein bottle coefficient has to satisfy~\cite{descendants}\footnote{We assume that the entries of the modular invariant $Z$ are $0$ or $1$. This
should always be the case if fixed points are correctly resolved.}  
\begin{equation} \label{eq:k}
	K^i=\epsilon_i Z_{ii} \;\; ,
\end{equation}
where  $\epsilon_i=\pm1$ and we define $\epsilon_i=0$ when $Z_{ii}=0$. This condition ensures that the sum $(T+K)/2$ contains sensible state multiplicities. Which Klein bottle coefficients $K^i$ are allowed? We have to require that in the new, unoriented theory, there is a zero probability of producing {\em unphysical} states, that is, states that are no longer present after the projection. For example, consider the case where we choose to symmetrize sectors $i$ and $j$. The symmetry of a sector is conserved in fusion: two states of opposite symmetry couple to an antisymmetric state, two states of equal symmetry couple to a symmetric state. So, when $k$ has a nonzero fusion with $i$ and $j$, we also have to symmetrize $k$. This can be summarized as follows: a Klein bottle projection $K^i=\epsilon_i Z_{ii}$ is consistent if and only if~\cite{descendants}
\begin{equation} \label{eq:consist}
	N_{ij}^{\;\;\;k} \epsilon_i  \epsilon_j \epsilon_{k} \geq 0 \;\; ,
\end{equation} 
for all $i,j$ and $k$; where $N_{ij}^{\;\;k}$ are the fusion coefficients given by the Verlinde formula \cite{verlinde}.  In section~\ref{sec-simple}, we describe a natural and simple way to find different projections. 

The next step in the construction is the following: one observes that the Klein bottle can have two physically different interpretations, called `channels'. In the direct channel, closed strings propagate in a loop while undergoing an orientation twist. This is essentially the partition function that we have discussed above.  The second channel is known as the transverse channel and is interpreted as the propagation of closed strings between two crosscap states. The transverse Klein bottle is
\begin{equation} \label{eq:transkb}
	\tilde{K} = \sum_i \Gamma_i^2 \chi_i \;\; .
\end{equation}

It is well known that both channels are related by a channel transformation, which in the case of the Klein bottle is the modular transformation $S$. Therefore, the coefficient $\Gamma_i$ and the Klein bottle coefficient $K^i$ are related:
\begin{equation} \label{eq:verlindeK}
	 K^i = \sum_m S_{m}^{~i} \Gamma_m \Gamma_m \;\; .
\end{equation}
for all $i$. 

We will now discuss the open sector. The most important diagram is the annulus: it relates the unknown open sector to the known closed sector by a channel transformation. In the transverse channel, closed strings propagate between two boundaries\footnote{Throughout this letter, we assume that the boundaries 
and the crosscaps 
do not break any of the worldsheet symmetry (`trivial gluing'). We also ignore Chan-Paton multiplicities and tadpole cancellation here, but they can easily be taken into account. We briefly comment on these subjects at the end of section~\ref{sec-simple}.} of opposite orientation. The amplitude is given by
\begin{equation} \label{eq:transA}
	\tilde{A}_{ab} = \sum_i B_{ia} B_{ib} \chi_i \;\; .
\end{equation}
The index $a$ labels the inequivalent boundaries.
 
The direct channel has the interpretation of an open string propagating in a loop and is therefore the open string partition function. It is given by
\begin{equation}
	A_{ab} = \frac{1}{2} \sum_{i} A^{i}_{\;ab} \chi_{i} \;\; .
\end{equation}
The annulus coefficients $A^{i}_{\;ab}$ are nonnegative integers. They count the number of  open string states in sector $i$ with boundary conditions $a$ and $b$ that exist in the theory. Since both channels are related by a channel transformation, which is $S$ in the case of the annulus, the annulus coefficients can be expressed in terms of the boundary coefficients $B_{ia}$ as follows:
\begin{equation} \label{eq:verlindeA}
	A^{i}_{\;ab}= \sum_m S^{~i}_{m} B_{ma} B_{mb} \;\; ,\label{eq:AtoB}
\end{equation}
for all $i,a$ and $b$. 

The last step in the construction of open descendants is a projection of the open sector to unoriented open string states. This is done by adding the M\"obius strip partition function to that of the annulus. As for the other two surfaces discussed so far, the M\"obius strip has two channels that are related by a channel transformation. Let us start with the transverse channel, which is interpreted as closed string propagation between a crosscap and a boundary $a$ (or vice versa). Therefore, this amplitude can be written as the geometric mean of transverse Klein bottle and annulus :
\begin{equation} \label{eq:transM}
	\tilde{M}_a = \pm \sum_{i} \Gamma_i B_{ia} \hat{\chi}_i \;\; .
\end{equation} 

In the direct channel the M\"obius strip is given by
\begin{equation} \label{eq:dirM}
	M_a =  \pm \frac{1}{2} \sum_{i} M^i_{~a} \hat{\chi}_i \;\; ,
\end{equation}
where the `twists' $M^i_{~a}$ are related to the boundary and crosscap coefficients via the channel transformation, which is $P=\sqrt{T}ST^2S\sqrt{T}$ for the M\"obius strip. So
\begin{equation} \label{eq:verlindeM}
	M^{i}_{~a} = \sum_m B_{ma} \Gamma_m P^{~i}_{m}\;\; , \label{CARDY}
\end{equation}
for all $i$ and $a$.

The M\"obius coefficients have to satisfy an extra condition~\cite{descendants}, namely
\begin{equation} \label{eq:m}
	|M^i_{~a}| \leq A^i_{\;aa} \;\; {\rm and} \;\; M^i_{~a} = A^i_{\;aa} \;\; {\rm mod} \, 2 \;\; .
\end{equation}
This ensures that the open sector has nonnegative, integer state degeneracies in the sum $(A+M)/2$.

Furthermore, the annulus coefficients have to satisfy two polynomial equations, known as `completeness conditions' \cite{completeness}. We define the reflection coefficients as $R_{ia}=B_{ia}\sqrt{S_{i0}}$. It can be shown \cite{completeness} \cite{stanev} \cite{zuber} that the completeness conditions are satisfied if the reflection coefficients are such that
\begin{equation}
\sum_{i} R_{ia} R_{ib}^*= \delta_{ab}\;\;\;,\;\;\; \label{R1}\\
\sum_{a} R_{ia} R_{ja}^* = \delta_{ij}. \label{R2}
\end{equation}
In unpublished work of Stanev~\cite{stanev}, similar completeness conditions are derived for Klein bottle, M\"obius strip and annulus coefficients. These conditions are satisfied when (\ref{eq:verlindeK}), (\ref{eq:verlindeA}), (\ref{eq:verlindeM}) and (\ref{R1}) hold.

To summarize: the art of constructing open descendants is to find the correct annulus and M\"obius coefficients, $A^{i}_{\;ab}$ and $M^i_{~a}$, for a given modular invariant $Z_{ij}$ and consistent Klein bottle projection $K^i$. This amounts to solving equations~(\ref{eq:consist}),~(\ref{eq:verlindeK}),~(\ref{eq:verlindeA}),~(\ref{eq:verlindeM}),~(\ref{eq:m}) and~(\ref{R1}). All these equations can in principle be solved by brute force, but this is not very satisfying. We believe that, in order to unravel the secrets of open strings, we should try to find model-independent solutions.    

Such a solution is known when the modular invariant is the charge conjugation invariant $C$. In this case, the primary index set $i$ and boundary index set $a$ are the same. Cardy~\cite{cardy} has found an expression for the annulus coefficients. They are simply the fusion rule coefficients:
\begin{equation} \label{eq:cardyA}
	A^i_{\;ab} = N^i_{\;ab} \;\; .
\end{equation}

The other coefficients in the C-diagonal case can be expressed in terms of an integer-valued tensor~\cite{planar} 
\begin{equation} \label{eq:verlindeY}
	Y_{ij}^{\;\;\;k} = \sum_{m} \frac{S_{mi} P_{mj} P_{m}^k}{S_{m0}} \;\; .
\end{equation}
The M\"obius and Klein bottle coefficients are in this case~\cite{descendants}
\begin{eqnarray} \label{eq:cardyM}
 	M^i_{~a}  =  Y_{a0}^{\;\;\;i} \;\;\;,\;\;\;
	K^i  =  Y^i_{\;\;00} \;\; . \label{eq:cardyK}
\end{eqnarray}

In the appendix, we show that $Y^i_{\;\;00}$ is the Frobenius-Schur indicator $\nu_i$ of a rational conformal field theory~\cite{bantay}: $\nu_i=1$ when $i$ is a real representation, $\nu_i= -1$ for a pseudoreal representation and $\nu_i = 0$ for a complex representation. But this immediately implies
\begin{equation} \label{eq:delta}
	|Y^i_{\;\;00}| = \delta_{i,i^c} \;\; ,
\end{equation}
in agreement with equation~(\ref{eq:k}) with $Z=C$. The superscript $c$ denotes the charge conjugate of a field.

This observation also implies that  condition (\ref{eq:consist}) is
satisfied, provided that non-vanishing Frobenius-Schur indicators
in conformal field theory 
are preserved in fusion, as they are in Lie-algebra tensor products:
\begin{equation} \label{eq:nu}
	\nu_i \nu_j = \nu_k,\ \   \ \hbox{for~} N_{ij}^{\;\;\;k} \neq 0\ ,
\nu_i\not=0,\ \nu_j\not=0,\ \nu_k\not=0\ .
\end{equation}
This clearly implies (\ref{eq:consist}).  

The boundary and crosscap coefficients are in this case 
\begin{equation}
	B_{ia} = \frac{S_{ia}}{\sqrt{S_{i0}}} \;\;\;\;\; , \;\;\;\;  \Gamma_{i} = \frac{P_{i0}}{\sqrt{S_{i0}}} \;\; \label{CaRdY}.
\end{equation}
Equations~(\ref{eq:verlindeK}),~(\ref{eq:verlindeA}),~(\ref{eq:verlindeM}) now reduce to equations~(\ref{eq:verlindeY}) and the Verlinde formula.

In the next section we extend the C-diagonal case such that it includes different Klein bottle projections. 

\section{Non-standard Klein bottles} \label{sec-simple}

We begin by repeating some well-known properties of simple currents that are needed for our discussion. A detailed review can be found in~\cite{revsimple}. 

A field $J$ satisfying
\begin{equation}
	J \times i = i' \;\; ,
\end{equation}
for all $i$, is called a simple current. When $J$ is a simple current, then so is any power of $J$. Every CFT contains one trivial simple current, the identity $0$.  We can associate a charge $Q_J (i)$ of a field $i$ with repect to the simple current $J$ defined by
\begin{equation}
	Q_J (i) = h_J + h_i - h_{J \times i} \;\; {\rm mod} \,1 \;\; .
\label{eq:Qdefin}
\end{equation}
where $h_i$ denotes the conformal weight of $i$. The order of a simple current is the smallest integer $N$ for which $J^N=0$. The charge of a field with respect to the simple current $J^d$ is $dQ_J$ and the charge with respect to the identity $0\, {\rm mod}\, 1$. So it follows that $Q_J=k/N \, {\rm mod}\, 1$, where $k$ is an integer. This charge is conserved (modulo integers) in the fusion of two fields $i$ and $j$:
\begin{equation} \label{eq:charge}
	Q_J(i) + Q_J(j) = Q_J(k) \;\; {\rm mod} \, 1 \;\; ,
\end{equation}
when $N_{ij}^{\;\;\;k} \neq 0$. Taking exponentials on both sides we get 
\begin{equation} \label{eq:phase}
	e^{2\pi i Q_J (i)} e^{2\pi i Q_J (j) } = e^{2\pi i Q_J (k)} \;\; ,
\end{equation}
when $N_{ij}^{\;\;\;k} \neq 0$. 

\subsection{Order $2$ simple current Klein bottles} \label{sec-simple2}

To make the presentation in this letter as transparent as possible, we will for the moment restrict to (half-)integer spin, order $2$ simple currents. In the next subsection, we discuss order $N$ simple currents. Let $L$ be an order  $2$ simple current. Then $e^{2\pi i Q_L (i)} = \pm 1$ and we have arrived at a natural Klein bottle projection:
\begin{equation} \label{eq:ntkb}
	K_{[L]}^i = e^{2\pi  i Q_L(i)} Y^i_{\;\;00} = Y^i_{\;LL} \;\; .
\end{equation}
(We used equation~(\ref{eq:y2}) and the fact that $L$ is order $2$ and therefore real, to derive the last step in this expression.) From equation~(\ref{eq:phase}), ~(\ref{eq:nu}) and~(\ref{eq:delta}), it follows that this Klein bottle satisfies the conditions~(\ref{eq:k}) and~(\ref{eq:consist}). 

We now proceed by deriving the other coefficients. From equation~(\ref{eq:verlindeK}),~(\ref{eq:ntkb}) and~(\ref{eq:verlindeY}) we immediately see that the crosscap coefficient must 
be\footnote{One could allow signs in this relation. An overall sign only 
affects the sign of the M\"obius amplitude, and is taken into account later.
Typically signs that depend on $i$ destroy the relation between the
M\"obius amplitude and the tensor $Y$, and hence invalidate integrality
and positivity of the partition function. However, we cannot rule
out that such signs might be acceptable in some special cases.}  
\begin{equation} \label{eq:ntc}
	  \Gamma_{[L]i} = \frac{P_{iL}}{\sqrt{S_{i0}}} \;\; .
\end{equation}
Now we need to determine the boundary coefficients $B_{ai}$. Suppose for the moment 
that they do not change.   
Then, from~(\ref{eq:verlindeM}),~(\ref{eq:verlindeY}) and~(\ref{eq:ntc}), the M\"obius coefficient becomes
\begin{equation} \label{eq:ntm}
	M^i_{[L]a} = Y_{aL}^{\;\;\;i}  \;\; .
\end{equation}
From equation~(\ref{eq:fs4}) in the appendix, we see that this M\"obius strip coefficient satisfies
\begin{equation}
	|M^i_{[L]a}| \leq N^{Li}_{\;\;aa} \;\;\;\; {\rm and} \;\;\; M^i_{[L]a} = N^{Li}_{\;\;aa} \, {\rm mod} \, 2 \;\; .
\end{equation}   
By equation~(\ref{eq:m}), a natural candidate for the annulus coefficients therefore is
\begin{equation} \label{eq:nta}
	A^i_{[L]ab} = N^{Li}_{\;\;ab} \;\; .
\end{equation}
This may appear to be in contradiction with (\ref{eq:AtoB}) and the earlier assumption that
the boundary coefficients do not change, but we will see in a moment that
there is in fact no contradiction, at least when $L$ has (half-)integer spin and is of  order $2$. 

To conclude: for every (half-)integer spin, order $2$ simple current $L$, there is a class of open descendants of a C-diagonal closed oriented theory, whose amplitudes are given by
\begin{eqnarray}
	K_{[L]}  =  \frac{1}{2} \sum_i Y^i_{\;LL} \chi_i  
\;\;\; &,& \;\;\;	 	\tilde{K}_{[L]}  =  \sum_i \frac{P_{iL}P_{iL}}{S_{i0}} \chi_i \;\; , \label{eq:solK} \\
	A_{[L]ab}  =  \frac{1}{2} \sum_{i} N^{Li}_{\;\;ab} \,\chi_{i} \;\;\; &,& \;\;\; 
	\tilde{A}_{[L]ab}  =  \sum_{i} \frac{e^{2\pi i Q_L (i)} S_{ia} S_{ib}}{S_{i0}} \, \chi_i \;\; , \label{eq:tA1} \\
	M_{[L]a} =   \pm \frac{1}{2} \sum_{i} Y_{aL}^{\;\;\;i} \,\hat{\chi}_i \;\;\; &,& \;\;\; 
	\tilde{M}_{[L]a}  =  \pm \sum_{i} \frac{S_{ia} P_{iL}}{S_{i0}} \,\hat{\chi}_i \;\; . \label{eq:solM}
\end{eqnarray}
We note that these expressions agree with the amplitudes of some special cases displayed in~\cite{planar} and~\cite{properties}. 

Note that the transverse M\"obius strip is still the geometric mean of the transverse Klein bottle and annulus, despite of the appearance of the phase $e^{2\pi i Q_L (i)}$ in equation~(\ref{eq:tA1}). The reason is that one may describe all quantities as in 
formulas (\ref{eq:transkb})--(\ref{eq:verlindeM}), but 
with
modified boundary  and crosscap coefficients 
\begin{equation}
	B_{[L]ia}   =   {S_{ia}\over \sqrt{S_{iL}}}\ ,         \;\; 
\quad \Gamma_{[L]i} = \frac{P_{iL}}{\sqrt{S_{iL}}} . \label{eq:FINAL}
\end{equation}
This 
appears to introduce an extra phase in the crosscap coefficient in
comparison with (\ref{eq:ntc}),
but it may be shown that $P_{iL}$ vanishes precisely when $Q_L(i)\in Z+1/2$ and $L$ (half-)integer spin. 
On the other hand, the extra phase in the boundary coefficient yields 
precisely the modified annulus, but does not affect the M\"obius strip, again because $P_{iL}=0$ when this phase is nontrivial.
This is
the way out of the apparent contradiction alluded to above. Note however that the assumption we made that the boundary coefficients do not change, is wrong. They {\em do} change, but this change is insignificant for order $2$ (half-)integer spin simple currents, as we have seen.  In the next subsection, when we consider order $N$ simple currents, we will first assume that equation~(\ref{eq:FINAL}) gives the correct boundary coefficient and then give evidence for this ansatz.

\subsection{Order $N$ simple current Klein bottles} \label{sec-simpleN}

The form of equation (\ref{eq:FINAL}) suggests a natural generalization to higer order simple currents. Consider an order $N$ simple current $J$, i.e. $J^N=0$. The charges of the fields with respect to $J$ are multiples of $1/N$. Nevertheless, consider the Klein bottle
\begin{equation} \label{eq:ntkb2}
	K_{[J]}^i = e^{2\pi  i Q_J(i)} Y^i_{\;\;00} = Y^{i\;J}_{\;J} \;\; .
\end{equation}
We will now argue that this Klein bottle is consistent. An obvious problem is of course what to do with the fields that have non-(half-)integer charges, since these charges lead to unexceptable phases. However, since the charge of a field is minus the charge of its conjugate (modulo $1$), all fields with non-(half-)integer charges are complex, so the corresponding phases are killed by the Frobenius-Shur indicator, $\nu_i = Y^i_{\;\;00}$. Furthermore, equation~({\ref{eq:ntkb2}}) obviously satisfies equation~({\ref{eq:k}}) and~(\ref{eq:consist}), so equation~({\ref{eq:ntkb2}}) is a consistent Klein bottle.

What are the other coefficients? After some algebra (equation~(\ref{eq:p}) might be helpful), we find precisely the same crosscap coefficient as in equation~(\ref{eq:FINAL})\footnote{One may use any unambiguous definition of the square root in the complex plane. Note that in all relevant quantities one only encounters the square or the absolute value squared of $\sqrt{S_{iJ}}$.}:
\begin{equation}
	\Gamma_{[J]i} = \frac{P_{iJ}}{\sqrt{S_{iJ}}} \;\; . \label{GJ}
\end{equation}
From the last subsection, we have seen that the boundary coefficients also change (see equation~(\ref{eq:FINAL})). Let us take as an ansatz for the boundary coefficients
\begin{equation} \label{eq:FINAL2}
	B_{[J]ia}   =   {S_{ia}\over \sqrt{S_{iJ}}} \;\; .   \label{BJ} 
\end{equation}
We will now give evidence for the correctness of this ansatz. By equations~(\ref{eq:verlindeM}) and~(\ref{eq:verlindeA}), the M\"obius and annulus coefficients become
\begin{equation} \label{eq:FINALma}
		M^i_{[J]a} = Y_{J^ca,J}^{\;\;\;\;\;\;\;\;i}  \;\; , \;\; A^i_{[J]ab} = N^{Ji}_{\;\;\;ab} \;\; .
\end{equation}
From equations~(\ref{eq:fs4}) and~(\ref{eq:s}), it follows that these coefficients satisfy the positivity and integrality conditions~(\ref{eq:m}). It may also be shown that for $J$ order $2$ and (half-)integer spin, these coefficients reduce to those of the last subsection, as expected. It is straightforward to show that this set of coefficients respect (\ref{R1}) and hence also the completeness conditions of~\cite{completeness} and~\cite{stanev}. This is strong evidence that our ansatz for the boundary coefficient, equation~(\ref{eq:FINAL2}), is indeed correct.

To conclude: For every simple current $J$, there is a class of open descendants of a C-diagonal closed oriented theory, whose amplitudes are given by
\begin{eqnarray}
	K_{[J]}  =  \frac{1}{2} \sum_i Y^{i\;\;J}_{\;J} \chi_i \;\;\; &,& \label{eq:solK2} \;\;\;
	\tilde{K}_{[J]}  =  \sum_i \frac{P_{iJ}P_{iJ}}{S_{iJ}} \chi_i \;\; , \\
	A_{[J]ab}  =  \frac{1}{2} \sum_{i} N^{Ji}_{\;\;\;ab} \,\chi_{i} \;\;\;&,&\;\;\; 
	\tilde{A}_{[J]ab}  =  \sum_{i} \frac{S_{ia} S_{ib}}{S_{iJ}} \, \chi_i \;\; , \label{eq:tA} \\
	M_{[J]a}  =   \pm \frac{1}{2} \sum_{i} Y_{J^ca,J}^{\;\;\;\;\;\;\;\;\;i} \,\hat{\chi}_i \;\;\; &,&\;\;\;
	\tilde{M}_{[J]a}  =  \pm \sum_{i} \frac{S_{ia} P_{iJ}}{S_{iJ}} \,\hat{\chi}_i \;\; . \label{eq:solM2}
\end{eqnarray}
We have argued that these amplitudes satisfy consistency of the Klein bottle projection, positivity and integrality of the open string sector and the completeness conditions.

It is in particular  instructive to consider the identity sector $i=0$, from which the
gauge bosons emerge. If we allow for non-trivial Chan-Paton multiplicities $n_a$ the 
number of massless vector bosons is
\begin{equation}
 {1\over2}\left[\sum_{ab} n_a n_b N^{J}_{~ab} \pm \sum_a n_a Y_{J^ca,J}^{~~~~~0}\right] . 
\end{equation}
In the standard case, $J=0$, this reduces to 
${1\over2} \sum_a (n_a n_{a^c} \pm n_a \nu_a) $,
where $\nu_a$ is the Frobenius-Schur indicator, and $a^c$ denotes the
charge conjugate of $a$. The overall sign of the M\"obius strip is fixed by tadpole cancellation 
to $-{\rm sign}(P_{00})$.
Given this sign, the type of gauge group one gets is completely
determined by the Frobenius-Schur indicator: $\nu_a=0$ implies $a\not=a^c$
and leads to $U(n_a)$ gauge groups (this requires $n_a=n_{a^c}$), whereas
$\nu_a=1 (-1)$ leads to $SO(n)$ ($Sp(n)$) if $P_{00}$ is positive (and
vice-versa if $P_{00}$ is negative). Analogous conclusions hold for
non-trivial Klein bottles, but in terms of a  ``generalized" Frobenius-Schur
indicator $Y_{J^ca,J}^{~~~~~0}$.

Equations (\ref{GJ}) and (\ref{BJ}) summarize our results. In this form the simple current origin of modified Klein bottles is most transparent. In fact the only change one has to make in comparision with (\ref{CaRdY}) is to replace all occurences of the identity field $0$ by a simple current $J$, with no restriction on its spin or order.

It is in general not true that all simple currents lead to different string theories, but we have checked explicitly that many of them do.

An obvious question one might ask is: can {\em all} nontrivial Klein bottles be obtained by simple currents? This question is unfortunately still open.

\appendix

\section{Some algebra with $J$'s, $P$'s and $Y$'s} \label{sec-form}

In this appendix, we present some useful formulas that are needed in the main text.  $J$ denotes an order $N$ simple current. We start by recalling~\cite{simple} that
\begin{equation} \label{eq:s}
	S_{Ji,j} = e^{2\pi i Q_{J}(j)} S_{ij} \;\;\; , \;\;\; T_{Ji} = e^{2\pi i [h_J-Q_J(i)]} \, T_i \;\; .
\end{equation} 
From this we can derive that
\begin{equation}
	P_{Ji,j}   =   e^{\pi i[\hat Q_J(j) - \hat Q_J(i)]} P_{i,Jj}         \;\; . \label{eq:p} \\
\end{equation}
The purpose of the hats on $Q$ is to avoid
sign ambiguities related
to the definition of the square root of $T$ in the
definition of $P$. Indeed, if we use
definition (\ref{eq:Qdefin}) for the charge, then (\ref{eq:p}) is only
defined up to a sign. This ambiguity is easily resolved by defining
$\sqrt{T} \equiv {\rm exp}[ \pi i (h-c/24)]$, and by defining $\hat Q$
as in (\ref{eq:Qdefin}) but {\it without} the ${\rm mod}~1$ freedom. 

With the use of equations~(\ref{eq:s}) and ~(\ref{eq:p}) we can calculate
\begin{equation}
	Y_{i,Jj}^{\;\;\;Jk} =   e^{\pi i [\hat{Q}_J(k) - \hat{Q}_J(j) - 2\hat{Q}_J (i)]}  Y_{ij}^{\;\;\;k} \label{eq:y1} \;\; .
\end{equation}
Taking $j=k=0$ in equation~(\ref{eq:y1}) gives
\begin{equation} \label{eq:y2}
	Y^{i\;\;J}_{\;J} = e^{2\pi i Q_J (i)}  Y^i_{\;\;00} \;\; ,
\end{equation}
Taking $i=a, j=0, k=J^c i$, equation~(\ref{eq:y1}) gives
\begin{equation} \label{eq:y3}
	|Y_{aJ}^{\;\;\;i}| = |Y_{a0}^{\;\;\;J^c i}| \;\; .
\end{equation}
We now prove that $Y_{i00}$ is the Frobenius-Schur indicator. Let us first rewrite the tensor $Y_{ijk}$ in the following way
\begin{eqnarray}
	Y_{ij}^{\;\;k} & = &  \sum_{m} \frac{S_{im} P_{jm} P^*_{km}}{S_{0m}} \nonumber \\ 
			& = &  \sqrt{\frac{T_j}{T_k}} \sum_{mnr} \frac{S_{im} S_{jn} T_n^2 S_{nm} S^*_{mr} T^{2*}_r S^*_{rk}}{S_{0m}} \nonumber \\ 
			& = &  \sqrt{\frac{T_j}{T_k}} \sum_{nr} N_{in}^{\;\;r} S_{jn} S^*_{kr} \frac{T_n^2}{T_r^2} \label{eq:y4} \;\; .
\end{eqnarray} 
Taking $j=k=0$ in equation~(\ref{eq:y4}) gives
\begin{equation} \label{eq:fs1}
	Y^i_{\;\;00}  =  \sum_{nr} N^i_{nr} S_{0n} S_{0r} \frac{T_n^2}{T_r^2}  \;\; ,
\end{equation} 
which is the Frobenius-Schur indicator $\nu_i$ of~\cite{bantay}. Taking $j=0$ in equation~(\ref{eq:y4}) gives
\begin{equation} \label{eq:fs2}
	Y_{i0}^{\;\;\;k}  =  \sqrt{\frac{T_0}{T_k}} \sum_{nr} N^i_{nr} S_{0n} S^*_{rk} \frac{T_n^2}{T_r^2}  \;\; ,
\end{equation} 
It is shown in~\cite{bantay} that this object satisfies\footnote{The 
complex conjugation on $S_{rk}$ is missing in \cite{bantay}. This is confirmed by the author of this reference.} 
\begin{equation} \label{eq:fs3}
	|Y_{i0}^{\;\;k} | \leq N^k_{\;ii} \;\;\;\; {\rm and} \;\;\; Y_{i0}^{\;\;k} = N^k_{\;ii} \  {\rm mod} \, 2 \;\; .
\end{equation}   
We now combine equations~(\ref{eq:y3}) and~(\ref{eq:fs3}):
\begin{equation} \label{eq:fs4}
	|Y_{aJ}^{\;\;\;i} | \leq N^{J^ci}_{\;\;\;aa} \;\;\;\; {\rm and} \;\;\; Y_{aJ}^{\;\;\;i} = N^{J^ci}_{\;\;\;aa} \  {\rm mod} \, 2 \;\; .
\end{equation}   
This implies posivity and integrality of the open string partition function.
Interestingly, in \cite{bantay} these properties of (\ref{eq:fs2}) 
were derived from 
the requirement of positivity and integrality of the dimension of eigenspaces
of certain braid matrices. These dimensions turn out to be equal to 
the multiplicities of open string ground states!    

Finally we remark that these properties of the $Y$-tensor 
may also be derived from the fusion
coefficients of permutation orbifolds, derived in \cite{borisov}. 
\vspace{10mm}
\begin{center}
{\bf Acknowledgements}
\end{center}
We would like to thank P. Bantay, J. Fuchs and C. Schweigert for electronic discussions on issues concerning this paper. 
N. S. would like to thank Funda\c{c}\~ao para a Ci\^encia e Tecnologia for financial support under the reference BD/13770/97. 
After finishing this work we realized that somewhat similar ideas (though not related to simple currents)
have been proposed in unpublished work of B. Pioline \cite{pioline}.
His A,K and M coefficients are however different from ours,
and for that reason do not satisfy in general the positivity and
integrality requirements.

\end{document}